\documentclass[aps,pra,epsfigure,twocolumn,superscriptaddress]{revtex4-2}%
\usepackage[colorlinks=true,linkcolor=blue,urlcolor=blue,citecolor=blue,pdfusetitle]{hyperref}
\usepackage[english]{babel}
\usepackage{amsmath}
\usepackage[caption=false]{subfig}
\usepackage{graphicx,epstopdf}
\usepackage[table,xcdraw,dvipsnames]{xcolor}
\usepackage{amsfonts}
\usepackage{bbm}
\usepackage{amssymb}
\usepackage{color}
\usepackage{latexsym}
\usepackage{physics}
\usepackage{times,txfonts}
\usepackage{orcidlink}

\begin{document}
	
	\title{Suppressing Self-Discharging of Quantum Batteries by Cavity Interactions}
	
	\author{Anass Jad\,\orcidlink{0009-0001-6392-3043}}
	\email{jadanass@gmail.com}
	\affiliation{Laboratory of R\&D in Engineering Sciences, Faculty of Sciences and Techniques Al-Hoceima,
		Abdelmalek Essaadi University, Tetouan, Morocco}
	
	\author{Abderrahim El Allati\,\orcidlink{0000-0002-2465-8515}}
	\email{eabderrahim@uae.ac.ma}
	\affiliation{Laboratory of R\&D in Engineering Sciences, Faculty of Sciences and Techniques Al-Hoceima,
		Abdelmalek Essaadi University, Tetouan, Morocco}
	
	\author{Mohammad B. Arjmandi\,\orcidlink{0000-0002-9222-6765}}
	\email{arjmandi94@gmail.com}
	\affiliation{Wilczek Quantum Center, Shanghai Institute for Advanced Studies,
		University of Science and Technology of China, Shanghai 201315, China}
	
	\begin{abstract}
		We analyse a two-cavity architecture, in which a lossy cavity hosting $N$ qubits is coherently coupled to an auxiliary cavity, as a resource for the storage phase of an open quantum battery at non-zero temperature. Within a local Lindblad treatment in the resonant configuration, we find that the inter-cavity coupling enhances the suppression of self-discharging across every initial preparation, battery size, and temperature we examine, with the protection degrading smoothly as the mean thermal occupation increases. For a single qubit, the energy-basis coherence of a pure superposition leads to better long-time retention than fully excited state, highlighting the beneficial role of quantum coherence in protecting stored energy against thermal degradation. For two-qubit batteries, Bell-state preparations exhibit enhanced long-time ergotropy retention compared with the fully excited state, while the inclusion of qubit-qubit interactions produces only a weak dependence on the interaction type and strength within the parameter regime considered. Extending the analysis to multi-qubit GHZ-charged batteries with all-to-all Heisenberg interactions, we find that the normalized retained ergotropy increases monotonically with the number of qubits. This behavior is consistent with the collective enhancement of the qubit-cavity coupling in the symmetric Dicke manifold, indicating that larger quantum batteries can benefit from improved protection against self-discharge. These findings establish cavity-assisted protection as a promising strategy for mitigating self-discharging and realizing of long-lived quantum batteries in experimentally accessible platforms.
	\end{abstract}
	
	\maketitle

	\section{Introduction}
	\label{sec:introduction}
	
	A fully thermalized quantum battery contains no extractable work under unitary operations \cite{Pusz1978,Lenard1978,Allahverdyan2004}. The maximum work that can be extracted from a state via unitaries is quantified by its ergotropy, which vanishes for a thermal state. Preserving ergotropy over finite times therefore requires engineered environments, tailored system--bath couplings, or controlled dynamics that compete with the dissipative drift toward equilibrium~\cite{Carrega2020,Santos2021,GarciaPintos2020,Kamin2020,Farina2019,Arjmandi2022}. The practical question is not whether self-discharging due to energy leakage into the surrounding environment can be eliminated, but whether it can be controlled by a small number of accessible parameters, and whether such control persists as the number of battery cells grows~\cite{AlickiFannes2013,VinjanampathyAnders2016,Campaioli2018Review,Campaioli2024Review,QuachReview2023}.
	
	Strategies to suppress self-discharging can be divided into time-dependent control schemes, such as Floquet engineering~\cite{BaiAn2020} and measurement-based feedback~\cite{MitchisonGooldPrior2021,YaoShao2022}, and bath-engineering schemes, which convert dissipation into a charging resource~\cite{Barra2019,Kamin2024Ancillas,Lemou2026,Khoudiri2026,Tabesh2020}, exploit disorder and many-body effects~\cite{Arjmandi2022,Arjmandi2026}, or reshape the spectral density seen by the system through coupling to additional modes~\cite{Xu2024,Imara2026,Hadipour2024,Xu2023,Cavaliere2025}. 
	
	The work reported here belongs to the latter category, for which we take as a starting point the minimal two-cavity architecture, proposed in Ref.\cite{Man2015}, in which coherent coupling between two lossy cavities renders the reduced dynamics of the embedded qubits non-Markovian. The mechanism we examine is therefore a cavity-induced modification of the system--bath dynamics, controlled by the inter-cavity coupling. The architecture has been studied in connection with the inhibition of self-discharging under non-Markovian noise~\cite{Xu2024} and, in a related collective setting, in Dicke and Tavis--Cummings batteries with collective dissipation~\cite{Canzio2025,PokhrelGea2025}. Existing analyses, however, have focused on the zero-temperature limit, whereas realistic cavity- or circuit-QED implementation operates at a finite temperature at which the thermal occupation of the cavity modes is non-negligible and feeds the dissipative dynamics~\cite{Blais2021cQED}. Moreover, the collective advantages established during the charging process~\cite{Ferraro2018,Khoudiri2025,Crescente2020,Campaioli2017,JadElAllati2026} do not necessarily extend to the self-discharging phase, whose dynamics is governed by dissipation rather than coherent driving.
	
	We address the aformentioned points  using a model of identical two-level systems embedded in a lossy cavity coherently coupled, through a photon-exchange
	coupling of strength $J$, to an auxiliary cavity containing no qubits. Each cavity undergoes independent losses into its own thermal bath within a local Lindblad framework whose parameter window lies in the regime in which local and global treatments are known to agree~\cite{Hofer2017,Cattaneo2019,DeChiara2018}. The principal results are as follows.
	
	First, the  inter-cavity coupling controls the long-time retention of the ergotropy across every preparation, battery size, and temperature we examine, with the protection degrading smoothly as the mean thermal occupation number increases. Remarkably, despite the additional dissipative processes introduced by thermal absorption, the long-time behavior remains primarily determined by the inter-cavity coupling.
	
	Second, coherent initial states retain a larger fraction of their ergotropy than incoherent preparations with the same initial extractable work. This asymmetry originates from the fact that thermal absorption affects qubit populations but does not generate energy-basis coherence, allowing coherence-assisted contributions to ergotropy to remain comparatively robust.
	
	Third, for GHZ-charged batteries of of more than two qubits coupled by an all-to-all Heisenberg interaction~\cite{Luo2025XYZ}, the normalised retained ergotropy at long times increases monotonically with the number of cells at every inter-cavity coupling explored. We attribute this dependence to the collective enhancement of the Tavis--Cummings coupling within the symmetric Dicke manifold, which redirects a growing fraction of the spontaneous emission into the mode protected by the auxiliary cavity. This provides a storage-phase analogue of the collective speed-up reported for the charging phase of quantum batteries~\cite{Ferraro2018,Crescente2020,Campaioli2017,JadElAllati2026}, with a distinct microscopic origin: the enhancement here protects stored work against dissipation rather than accelerating its accumulation.
	
	The remainder of the paper is organised as follows. Section~\ref{sec:model} specifies the model, the master equation, and the figure of merit. Section~\ref{sec:N1} treats the single-cell case. Section~\ref{sec:N2} extends the analysis to two qubits, with and without direct qubit--qubit interactions, and examines the sensitivity of the protection to a non-zero auxiliary-cavity loss rate. Section~\ref{sec:multi-qubit} analyses the dependence on the number of cells for GHZ-charged batteries and discusses experimental feasibility in circuit-QED. Section~\ref{sec:conclusion} concludes the paper.

	\section{Model}
	\label{sec:model}
	
\begin{figure}[htbp]
		\centering
		\includegraphics[width=\linewidth]{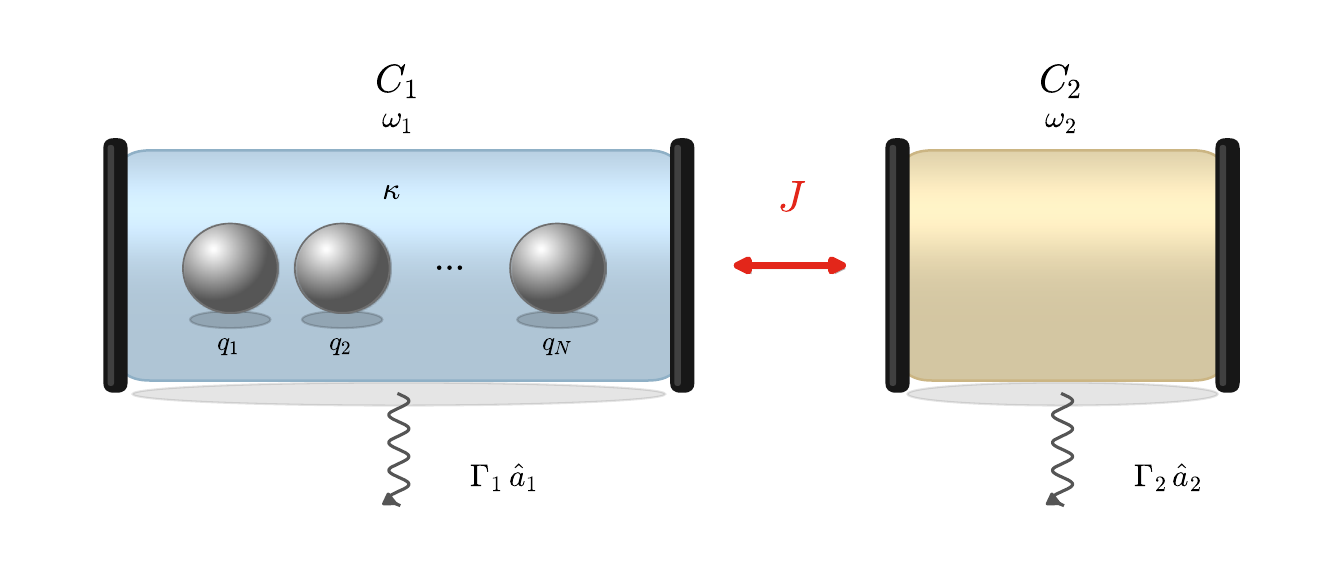}
		\caption{Sketch of the cavity-based quantum-battery architecture. A
			collection of $N$ identical qubits $\{q_1, q_2, \ldots, q_N\}$ of transition
			frequency $\omega_0$ is embedded in a common lossy cavity $C_1$ (frequency
			$\omega_1$, decay rate $\Gamma_1$) with which each qubit interacts at the same
			coupling strength $\kappa$. The cavity $C_1$ is coherently coupled to an
			auxiliary cavity $C_2$ (frequency $\omega_2$, decay rate $\Gamma_2$) through a
			photon-exchange coupling of strength $J$. Both cavities are initially in the
			vacuum state and dissipate independently into their own thermal reservoirs at
			temperatures $T_1$ and $T_2$ (wavy arrows), with mean photon occupations
			$\bar n_1$ and $\bar n_2$. The inter-cavity coupling $J$ and the number of
			qubits $N$ are the central control parameters of the self-discharging
			suppression studied in this work: $J$ sets the dark-mode protection, while
			increasing $N$ enhances the collective protection of the stored ergotropy.}
		\label{fig:scheme_multi}
	\end{figure}

	We consider a battery consisting of  $N$ identical two-level systems of transition frequency $\omega_0$, embedded in a single lossy cavity $C_1$ of frequency $\omega_1$ and photon decay rate $\Gamma_1$. The cavity $C_1$ is coupled, through a photon-exchange
	coupling of strength  $J$, to an auxiliary cavity $C_2$ of frequency $\omega_2$ and decay rate $\Gamma_2$. The qubits do not couple directly to a thermal bath; their relaxation proceeds through the common mode of $C_1$. Direct qubit relaxation channels are neglected on the assumption that Purcell-enhanced decay through $C_1$ dominates intrinsic qubit decoherence, as is generically the case in cavity-QED implementations of this architecture~\cite{Blais2021cQED}. A schematic representation is given in Fig.~\ref{fig:scheme_multi}.
	
	\subsection{Hamiltonian and master equation}
	\label{sec:model:H}
	
	We work in units where $\hbar = 1$ and $k_B = 1$, and use $\Gamma_1$ as the frequency scale, so that time is measured in units of $\Gamma_1^{-1}$. Within the rotating-wave approximation, the total Hamiltonian decomposes as
	\begin{equation}
		\hat H \;=\; \hat H_B + \hat H_C + \hat H_{BC} + \hat H_{CC},
		\label{eq:H_decomp}
	\end{equation}
	with
	\begin{subequations}
		\label{eq:H_terms}
		\begin{align}
			\hat H_B &= \hat H_0^{B} + \hat H_{\mathrm{int}}^{B},
			\qquad
			\hat H_0^{B} = \frac{\omega_0}{2}\sum_{i=1}^{N}\hat\sigma_z^{(i)},
			\label{eq:HB_def}\\
			\hat H_C &= \omega_1\,\hat a_1^\dagger\hat a_1 + \omega_2\,\hat a_2^\dagger\hat a_2,
			\label{eq:HC_def}\\
			\hat H_{BC} &= \kappa\sum_{i=1}^{N}\bigl(\hat a_1^\dagger\hat\sigma_-^{(i)} + \hat a_1\hat\sigma_+^{(i)}\bigr),
			\label{eq:HBC_def}\\
			\hat H_{CC} &= J\bigl(\hat a_1^\dagger\hat a_2 + \hat a_1\hat a_2^\dagger\bigr).
			\label{eq:HCC_def}
		\end{align}
	\end{subequations}
	Here $\hat\sigma_z^{(i)}$ and $\hat\sigma_\pm^{(i)}$ are the Pauli and ladder operators of the $i$-th qubit, and $\hat a_n$, $\hat a_n^\dagger$ are the annihilation and creation operators of cavity $C_n$. The term $\hat H_{BC}$ is the standard Tavis--Cummings coupling~\cite{TavisCummings1968}, $\hat H_{CC}$ is the linear photon-exchange interaction between $C_1$ and $C_2$, and $\hat H_{\mathrm{int}}^{B}$ is a direct qubit--qubit interaction that we leave unspecified at this stage; it will be defined in Secs.~\ref{sec:N2}--\ref{sec:multi-qubit}, and is set to zero in the absence of direct couplings (Sec.~\ref{sec:N1}).
	
	The model is parametrised by $\omega_0$, $\omega_{1,2}$, $\kappa$, $J$, and $\Gamma_{1,2}$. Throughout Secs.~\ref{sec:N1}--\ref{sec:multi-qubit} we work in the resonant condition $\omega_0 = \omega_1 = \omega_2 \equiv \omega$.

	Each cavity is weakly coupled ($\Gamma_n \ll \omega_n$) to its own bosonic bath in thermal equilibrium at temperature $T_n$, with mean occupation $\bar n_n = (e^{\omega_n/T_n} - 1)^{-1}$. The system--bath dynamics is described by the Lindblad master equation~\cite{BreuerPetruccione,GardinerZoller}
	\begin{equation}
		\dot\rho = -i[\hat H,\rho]
		+ \sum_{n=1,2}\Bigl\{
		\Gamma_n(\bar n_n + 1)\,\mathcal{D}[\hat a_n]\rho
		+ \Gamma_n\bar n_n\,\mathcal{D}[\hat a_n^\dagger]\rho
		\Bigr\},
		\label{eq:master}
	\end{equation}
	with $\mathcal{D}[\hat X]\rho = \hat X\rho\hat X^\dagger - \tfrac{1}{2}\{\hat X^\dagger\hat X,\rho\}$. The two dissipators in Eq.~\eqref{eq:master} account for spontaneous emission and thermally-induced absorption, whose rates satisfy the detailed-balance ratio $\Gamma_n\bar n_n/[\Gamma_n(\bar n_n+1)] = e^{-\omega_n/T_n}$. This guarantees that an isolated cavity ($\kappa = J = 0$) relaxes to its Gibbs state at temperature $T_n$. The zero-temperature limit $\bar n_n \to 0$ recovers the single-channel dissipator used in Ref.~\cite{Man2015}.
	
	The Lindblad operators in Eq.~\eqref{eq:master} act on the bare cavity modes rather than on the normal modes of the coupled-cavity Hamiltonian. This local form of the dissipator is appropriate for the parameter window considered here, in which local and global treatments are known to agree ~\cite{Hofer2017,Cattaneo2019,DeChiara2018} the correspondence of these dimensionless parameters to laboratory units is discussed in (Sec.~\ref{sec:experimental}).
	
	The battery is initially prepared in a charged $N$-qubit state $\rho_B(0)$, while both cavities are in the vacuum, so that $\rho(0) = \rho_B(0)\otimes |0_{C_1},0_{C_2}\rangle\langle 0_{C_1},0_{C_2}|
	$. The reduced state of the battery, $\rho_B(t) = \Tr_{C_1 C_2}[\rho(t)]$, is the object whose energetic properties we analyse.

	\subsection{Figure of merit and control parameters}
	\label{sec:model:ergotropy}
	
	The figure of merit is the ergotropy~\cite{Allahverdyan2004}, defined as the maximum work that can be unitarily extracted from $\rho_B$ with respect to its Hamiltonian $\hat H_B$:
	\begin{equation}
		\mathcal{W}(\rho_B)
		= \Tr(\rho_B\hat H_B)
		- \min_U\Tr(U\rho_B U^\dagger\hat H_B),
		\label{eq:ergotropy_def}
	\end{equation}
	where the minimisation runs over all unitaries on the battery Hilbert space of dimension $d_B = 2^N$. The minimum is reached by the unitary mapping $\rho_B$ to its passive state, that is, the state of the same spectrum as $\rho_B$ whose energy cannot be lowered by any unitary. This yields the closed-form expression \cite{Pusz1978,Lenard1978}
	\begin{equation}
		\mathcal{W}(\rho_B)
		= \Tr(\rho_B\hat H_B)
		- \sum_{k=1}^{d_B} r_k\,\varepsilon_k
		\label{eq:ergotropy_eigen}.
	\end{equation}
	 Where $\{r_k\}$ and $\{\varepsilon_k\}$ are the eigenvalues of $\rho_B$ and $\hat H_B$ respectively, ordered as $r_k \ge r_{k+1}$ and $\varepsilon_k \le \varepsilon_{k+1}$.
	
	Three dimensionless ratios characterise the regime of operation: $\kappa/\Gamma_1$ (qubit--cavity coupling versus cavity loss), $J/\Gamma_1$ (inter-cavity coupling versus cavity loss, the central control parameter of this work), and the mean thermal occupation $\bar n$, equivalently the dimensionless inverse temperature $\beta\omega$. Throughout this work we fix $\kappa = 0.24\,\Gamma_1$, which places the $J=0$, $\bar n=0$ limit inside the weak-coupling Markovian regime $\kappa \lesssim \Gamma_1/4$ of Ref.~\cite{Man2015}. This choice provides a stringent test for the proposed scheme: any retention observed at finite $J$ is then due to the structured environment generated by $C_2$.

	\section{Single-qubit case}
	\label{sec:N1}
	We begin with the single-qubit case, the simplest setting in which the mechanism
	underlying the cavity-mediated protection can be exhibited in closed form. In an
	open battery the stored work is lost to self-discharging as the qubit excitation
	decays through the lossy cavity $C_1$. Retaining the ergotropy therefore requires
	part of the excitation to be shielded from the corresponding loss channel
	$\mathcal{D}[\hat{a}_1]$. This shielding can stem from a dark mode of the
	coherent qubit--cavity dynamics, whose overlap with the charged state fixes the
	fraction of ergotropy retained at long times and is itself controlled by the
	inter-cavity coupling $J$. We introduce this mode below and use it throughout the
	section to interpret the thermal results; its derivation and its dependence on
	$J/\kappa$ are given in Appendix~\ref{app:darkmode}.
	
	Within the single-excitation manifold of qubit$+C_1+C_2$, the coherent Hamiltonian admits a dark mode
	\begin{equation}
		\ket{\psi_-} \;=\; \frac{1}{\sqrt{J^2+\kappa^2}}
		\bigl(J\,\ket{e,0,0} - \kappa\,\ket{g,0,1}\bigr),
		\label{eq:darkmode}
	\end{equation}
	annihilated by $\hat a_1$ and therefore stationary under $\mathcal{D}[\hat a_1]$ . For $\bar n = 0$ and $\Gamma_2 = 0$, an initial state with non-zero projection onto $\ket{\psi_-}$ retains the corresponding qubit coherence at long times, asymptotically proportional to a fraction $J^2/(J^2+\kappa^2)$ of its initial value~\cite{Man2015}. The present section examines the ergotropy of the reduced qubit state when both cavities are coupled to thermal reservoirs of mean occupation $\bar n > 0$, so that the excitation conservation underlying the dark-mode argument is broken by $\mathcal{D}[\hat a_n^\dagger]$.
	
We compare two initial preparations, with the ergotropy $\mathcal{W}_0$
evaluated with respect to the single-qubit battery Hamiltonian
$\hat H_B = \tfrac{\omega_0}{2}\,\hat\sigma_z$,
\begin{subequations}
	\label{eq:N1_states}
	\begin{align}
		\rho_B^{(1)}(0) &= \ketbra{e}{e},
		& \mathcal{W}_0 &= \omega_0,
		\label{eq:N1_state1}\\
		\rho_B^{(2)}(0) &= \ketbra{\psi_+}{\psi_+},
		& \mathcal{W}_0 &= \omega_0/2,
		\label{eq:N1_state2}
	\end{align}
\end{subequations}
	with $\ket{\psi_+} = (\ket{e}+\ket{g})/\sqrt{2}$. The preparation $\rho_B^{(1)}$ is fully excited, diagonal in the qubit energy eigenbasis, and stores its initial ergotropy entirely as a population imbalance. The preparation $\rho_B^{(2)}$ stores half its initial ergotropy as energy-basis coherence; the state $\ket{\psi_+}\!\ket{0,0}$ has overlap $J/[\sqrt{2}\sqrt{J^2+\kappa^2}]$ with the dark mode of Eq.~\eqref{eq:darkmode}.
	
	\begin{figure*}[t]
		\centering
		\includegraphics[width=\textwidth]{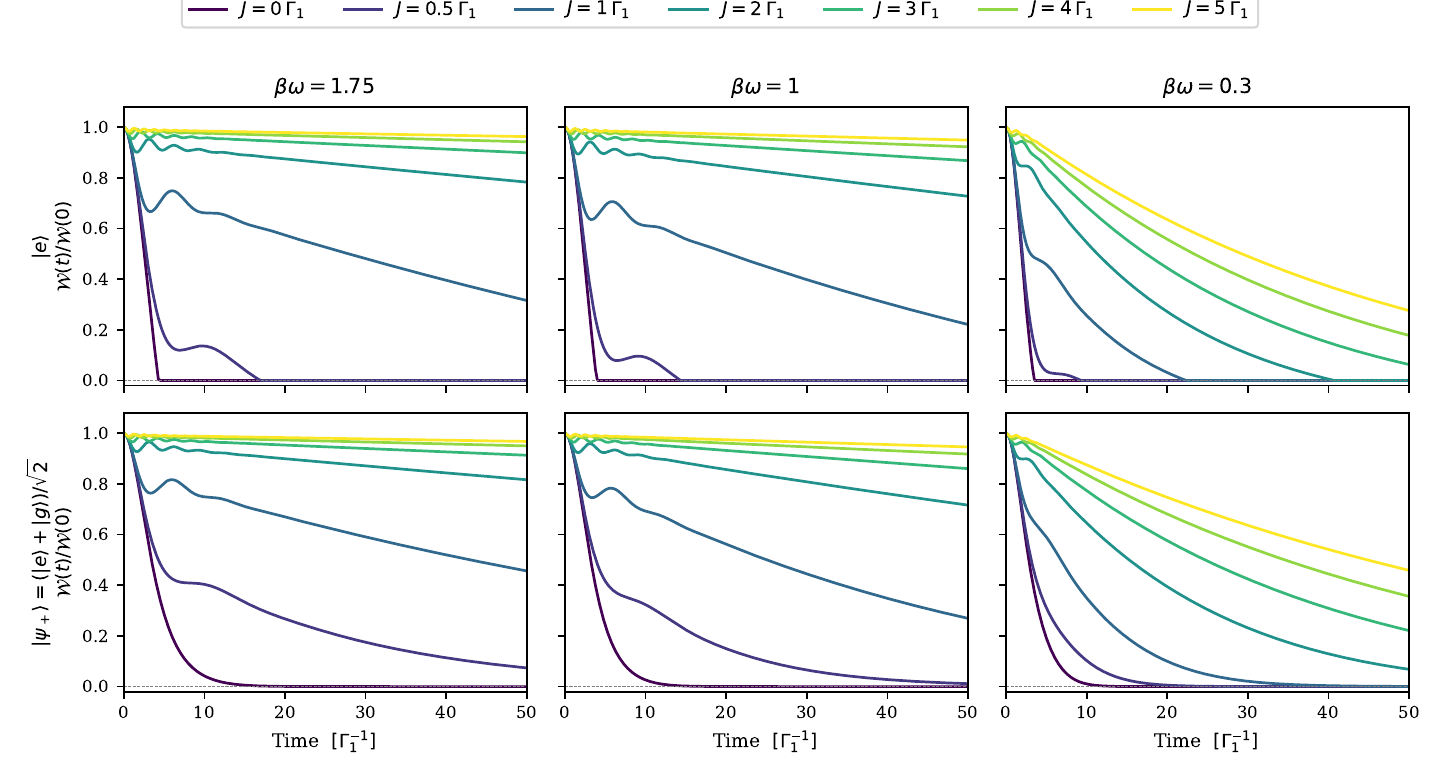}
	\caption{Normalised ergotropy $\mathcal{W}(t)/\mathcal{W}(0)$ of the single-qubit battery for the two initial states of Eqs.~\eqref{eq:N1_state1}--\eqref{eq:N1_state2} (rows; top: $\ket{e}$, bottom: $\ket{\psi_+}=(\ket{e}+\ket{g})/\sqrt{2}$), for seven values of the inter-cavity coupling $J/\Gamma_1 \in \{0,\,0.5,\,1,\,2,\,3,\,4,\,5\}$ (colour-coded from dark to light). Columns correspond to dimensionless inverse temperatures $\beta\omega \in \{1.75,\,1,\,0.3\}$, with both reservoirs at the same mean thermal occupation $\bar n_1=\bar n_2=\bar n = (e^{\beta\omega}-1)^{-1} \approx 0.21,\,0.58,\,2.88$. Parameters: $N=1$, $\kappa = 0.24\,\Gamma_1$, $\Gamma_2 = 0.1\,\Gamma_1$, resonant configuration $\omega_0 = \omega_1 = \omega_2 \equiv \omega$.}
	\label{fig:N1}
	\end{figure*}
	
	Figure~\ref{fig:N1} shows $\mathcal{W}(t)/\mathcal{W}(0)$ across the three thermal regimes. We organise the discussion around three observations.
	
	\emph{(i) Dependence on the inter-cavity coupling $J$.}\;
	For both initial preparations and at all temperatures considered,
	increasing $J$ enhances the retention of ergotropy over the simulated
	time window, while the case of $J=0$ exhibits the fastest
	self-discharging. The oscillations visible for $J\sim\Gamma_1$ at lower
	temperatures originate from coherent excitation exchange between the
	qubit and the auxiliary cavity $C_2$, mediated by $C_1$. These
	oscillations remain visible at $\beta\omega=1$ and become progressively
	damped as the temperature increases. The behaviour of the $J=0$ baseline
	is consistent with the dark-state structure of Eq.~\eqref{eq:darkmode}:
	for $J=0$, the dark state reduces to $\ket{g,0,1}$, which has no overlap
	with the initial excitation $\ket{e,0,0}$. Consequently, the qubit
	excitation remains fully exposed to cavity losses through
	$\mathcal{D}[\hat a_1]$.
	
	\emph{(ii) Smooth degradation with temperature.}\;
	The retention of ergotropy decreases progressively as the thermal
	occupation $\bar n$ increases. At low temperature ($\beta\omega=1.75$),
	the curves for $J \gtrsim 2\Gamma_1$ approach a long-lived plateau that
	remains close to unity over the simulated time interval for both initial
	preparations. As the temperature increases, this plateau is gradually
	suppressed and replaced by a slow residual decay. At $\beta\omega=0.3$,
	no clear plateau is visible within the interval $t\Gamma_1\in[0,50]$,
	and the ergotropy decreases approximately monotonically. Nevertheless,
	increasing $J$ continues to provide substantial ergotropy protection.
	
	\emph{(iii) Dependence on the initial preparation.}\;
	For all values of $J>0$ and $\beta\omega$ considered, the normalized
	ergotropy associated with the coherent superposition state
	$\ket{\psi_+}$ remains above that of the excited state $\ket{e}$. The
	difference is modest at low temperature ($\beta\omega=1.75$),
	particularly for $J\gtrsim2\Gamma_1$, and becomes progressively more
	pronounced as the thermal occupation increases. The largest separation
	is observed at $\beta\omega=0.3$, where the coherent preparation
	consistently retains a greater fraction of its initial ergotropy
	throughout the simulated time window.
	
	The observation~(iii) reflects how the two preparations
	store their work. The thermal absorption channels $\mathcal{D}[\hat
	a_n^\dagger]$ inject excitations into the cavities incoherently, and the
	Tavis--Cummings coupling transfers this to the qubit as population only.
	The work stored as energy-basis coherence in $\ket{\psi_+}$ is therefore
	not directly fed by the thermal flux, and is depleted only through the
	slower coherent qubit--cavity dynamics that $C_2$ suppresses via
	Eq.~\eqref{eq:darkmode}. The population imbalance of $\ket{e}$, by
	contrast, is drained directly by the thermal flux through $C_1$. As
	$\bar n$ grows, this flux increases, widening the gap between the two
	preparations.
	
	Section~\ref{sec:N2} examines whether the ordering between diagonal and coherent preparations persists at $N = 2$, and how it is affected by direct qubit--qubit interactions and by entangled preparations.

	\section{Two-qubit case}
	\label{sec:N2}
	
	We extend the analysis to the two-qubit case, $N=2$, which is the simplest setting in which genuine quantum correlations between battery cells and direct qubit--qubit interactions can be studied. All parameters are inherited from Sec.~\ref{sec:model}: $\kappa = 0.24\,\Gamma_1$, resonant configuration $\omega_0 = \omega_1 = \omega_2 \equiv \omega$.

	We consider three cases of preparation as:
	\begin{subequations}
		\label{eq:N2_states}
		\begin{align}
			\ket{\Psi_1} &= \ket{ee},
			\label{eq:N2_state1}\\
			\ket{\Psi_2} &= \bigl[(\ket{e} + \ket{g})/\sqrt{2}\bigr]^{\otimes 2},
			\label{eq:N2_state2}\\
			\ket{\Psi_3} &= \ket{\Phi^+}
			= \tfrac{1}{\sqrt{2}}\bigl(\ket{ee} + \ket{gg}\bigr),
			\label{eq:N2_state3}
		\end{align}
	\end{subequations}
including a fully excited diagonal state, a separable state with local coherence, and a maximally entangled Bell state, respectively. A weak Heisenberg coupling is included throughout this subsection to model the direct exchange generic in cQED implementations,
\begin{equation}
	\hat H_{\mathrm{int}}^{B}
	= g\bigl(\hat\sigma_x^{(1)}\hat\sigma_x^{(2)}
	+ \hat\sigma_y^{(1)}\hat\sigma_y^{(2)}
	+ \hat\sigma_z^{(1)}\hat\sigma_z^{(2)}\bigr),
	\label{eq:Hint_N2}
\end{equation}
with $g$ the coupling strength. We sweep $J/\Gamma_1 \in \{0,\,0.5,\,1,\,2,\,3,\,4,\,5\}$ at three inverse temperatures $\beta\omega \in \{1.75,\,1,\,0.3\}$ ($\bar n \approx 0.21$, $0.58$, $2.88$), matching the range used at $N=1$.
	\begin{figure*}[t]
		\centering
		\includegraphics[width=\textwidth]{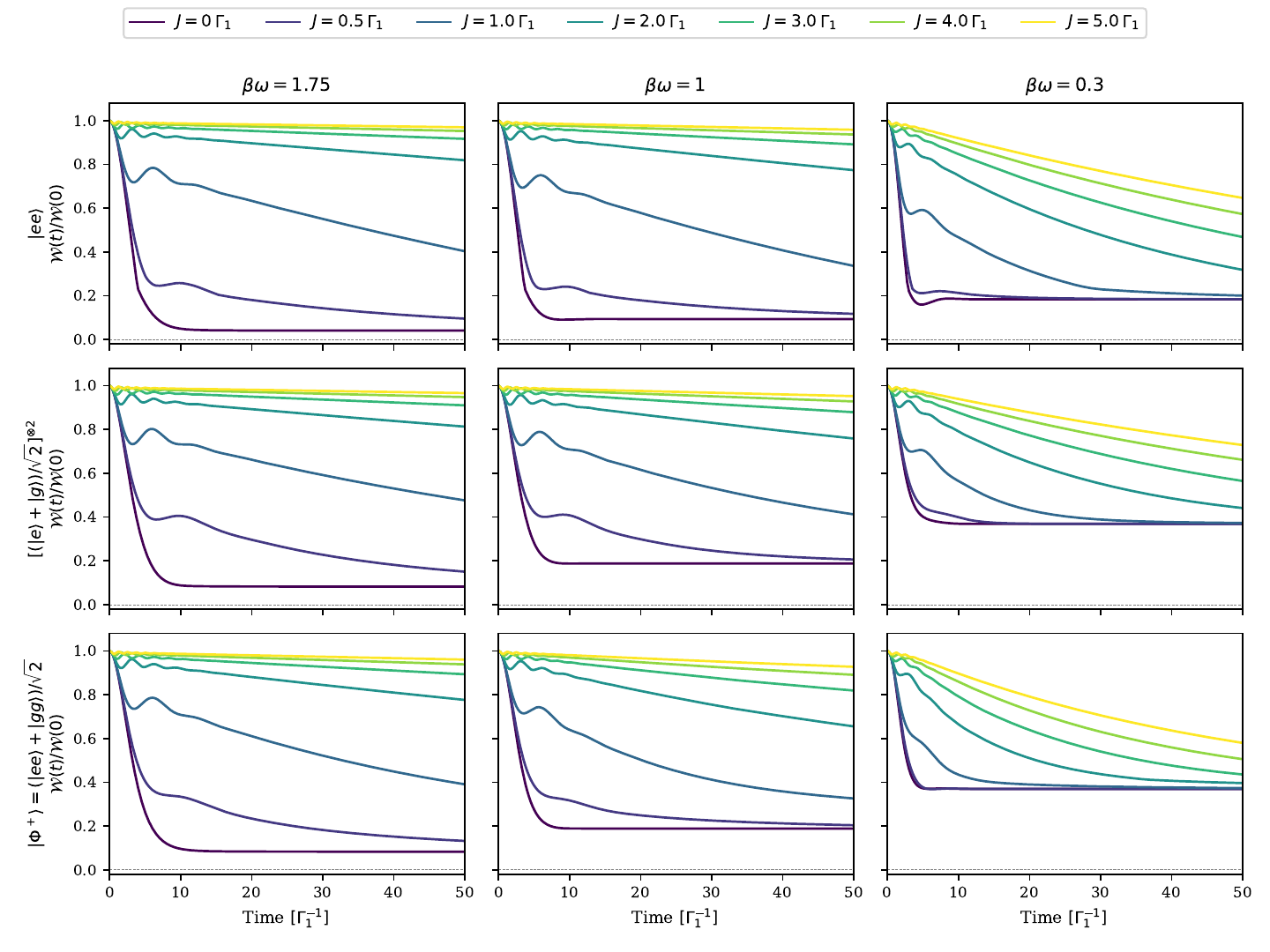}
		\caption{Normalised ergotropy $\mathcal{W}(t)/\mathcal{W}(0)$ of a two-qubit battery in cavity $C_1$ coupled to a  auxiliary cavity $C_2$, for three initial preparations (rows) and three inverse temperatures (columns). Top row: fully excited state $\ket{ee}$. Middle row: product superposition $[(\ket{e}+\ket{g})/\sqrt{2}]^{\otimes 2}$. Bottom row: Bell state $\ket{\Phi^+}=(\ket{ee}+\ket{gg})/\sqrt{2}$. Curves correspond to $J/\Gamma_1 \in \{0, 0.5, 1, 2, 3, 4, 5\}$ from dark to light. Parameters: $\omega_0=\omega_1=\omega_2=1$, $\kappa =0.24\,\Gamma_1$, $\Gamma_2=0.1\Gamma_1$, Heisenberg qubit--qubit coupling $g=0.1\,\Gamma_1$. for $\beta\omega \in \{1.75, 1, 0.3\}$ respectively.}
		\label{fig:N2_3states}
	\end{figure*}
	
	Figure~\ref{fig:N2_3states} shows that the qualitative trends identified
	in the single-qubit case persist for $N=2$. For all initial preparations
	and temperatures considered, increasing the inter-cavity coupling $J$
	improves ergotropy retention, while the $J=0$ configuration exhibits the
	fastest self-discharging. At low and intermediate temperatures
	($\beta\omega=1.75$ and $1$), the dynamics displays oscillations at
	intermediate values of $J$, reflecting coherent excitation exchange
	within the coupled qubit--cavity system. These oscillations become
	progressively damped as the thermal occupation increases.
	
	As in the single-qubit case, the protection mechanism weakens gradually
	as the temperature increases. For $\beta\omega=1.75$, couplings
	$J\gtrsim2\,\Gamma_1$ retain more than $90\%$ of the initial ergotropy at
	$t\Gamma_1=50$ for all preparations considered. At $\beta\omega=1$, a
	comparable level of protection is achieved only for the largest couplings
	studied. At $\beta\omega=0.3$, the ergotropy decreases more substantially
	over the simulated time window. Nevertheless, the beneficial effect of
	the auxiliary cavity remains evident, where larger values of $J$
	consistently yield higher long-time ergotropy, and a significant fraction
	of the initial ergotropy is still preserved at strong inter-cavity
	coupling.
	
	A feature specific to the two-qubit case emerges at finite temperature.
	For weak inter-cavity coupling ($J\lesssim0.5\,\Gamma_1$), the fully
	excited state $\ket{ee}$ relaxes to a lower long-time ergotropy than the
	states $\ket{\Psi_2}$ and $\ket{\Phi^+}$, which approach noticeably
	higher plateaus. The difference becomes more pronounced as the
	temperature increases. This behaviour is consistent with the trend
	already observed in the single-qubit case, namely that the retention of
	ergotropy depends not only on its initial magnitude but also on the
	structure of the initial state. In the parameter regime considered here,
	the correlated states $\ket{\Psi_2}$ and $\ket{\Phi^+}$ retain a larger
	fraction of their initial ergotropy than the fully excited product state
	$\ket{ee}$.
	
	Finally, for sufficiently strong protection ($J\gtrsim4\,\Gamma_1$) and
	moderate-to-low temperatures ($\beta\omega\ge1$), the differences between
	the three initial preparations become negligible on the scale of
	Fig.~\ref{fig:N2_3states}. In particular, the Bell state $\ket{\Phi^+}$
	exhibits no measurable advantage over the separable state $\ket{\Psi_2}$
	within the simulated time window.

	\begin{figure*}[t]
		\centering
		\includegraphics[width=\textwidth]{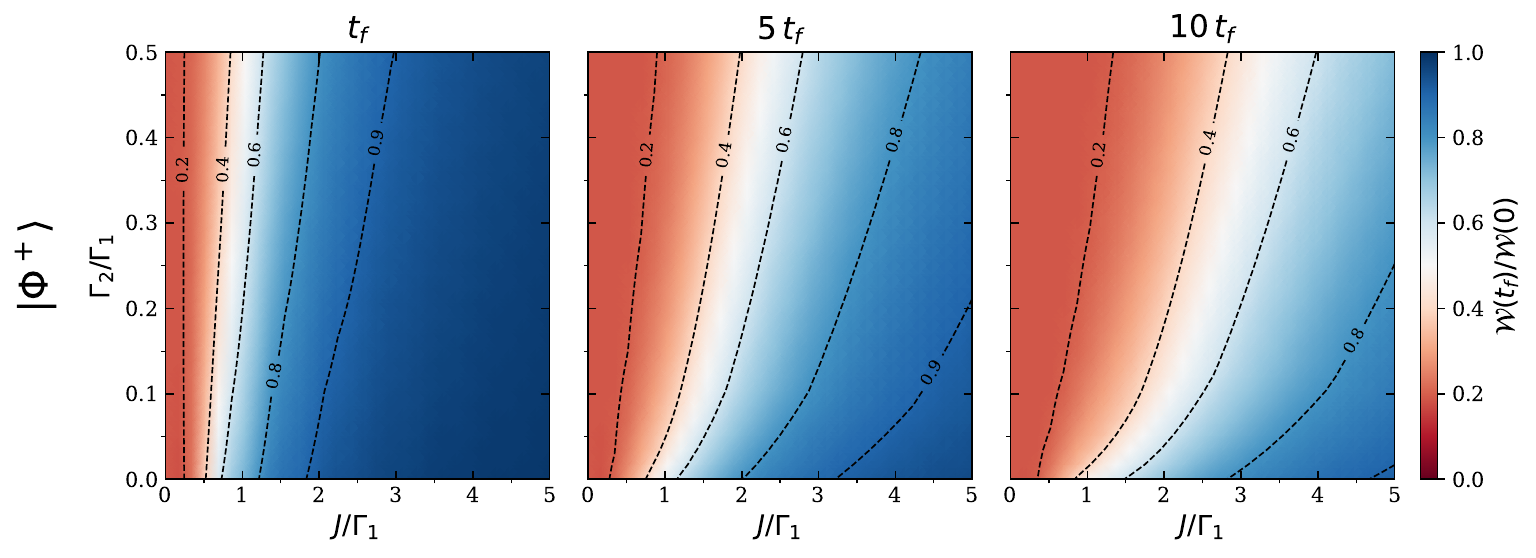}
		\caption{Robustness of the retained ergotropy $\mathcal{W}(t_f)/\mathcal{W}(0)$ for the Bell state $\ket{\Phi^+} = (\ket{ee} + \ket{gg})/\sqrt{2}$ in the plane of the inter-cavity coupling $J/\Gamma_1$ and the auxiliary loss rate $\Gamma_2/\Gamma_1$, at inverse temperature $\beta\omega = 1$. Panels show three evaluation times $t_f$, $5\,t_f$, and $10\,t_f$, with $t_f\Gamma_1 = 8.91$ chosen as the time at which the uncoupled ($J = 0$) baseline has relaxed to  $1\%$ of its initial amount of extractable energy. Dashed lines show contours of constant $\mathcal{W}(t_f)/\mathcal{W}(0) \in \{0.2, 0.4, 0.6, 0.8, 0.9\}$. Parameters: $\omega = 1$, $\kappa = 0.24\,\Gamma_1$, $g = 0.1\,\Gamma_1$.}
		\label{fig:N2_Phi}
	\end{figure*}
	
	Figure~\ref{fig:N2_Phi} shows the retained ergotropy
	$\mathcal{W}(t_f)/\mathcal{W}(0)$ as a function of the inter-cavity coupling
	$J/\Gamma_1$ and the auxiliary-cavity loss rate $\Gamma_2/\Gamma_1$ for three
	evaluation times. The results indicate that the retained ergotropy is governed
	primarily by the strength of the inter-cavity coupling. For weak coupling
	($J \lesssim \Gamma_1$), the contours are nearly insensitive to $\Gamma_2$.
	However, as $J$ increases, the dependence on $\Gamma_2$ becomes more apparent,
	with larger auxiliary losses reducing the retained ergotropy. Nevertheless, the
	variation induced by $\Gamma_2$ remains significantly smaller than that produced
	by changing $J$ over the same parameter range.
	
	This behaviour persists at $5t_f$ and $10t_f$. The results therefore suggest
	that the protection mechanism is controlled predominantly by the inter-cavity
	coupling, while remaining comparatively robust against moderate losses in the
	auxiliary cavity. In particular, the idealised choice $\Gamma_2 = 0$ captures
	the qualitative behaviour observed throughout the range
	$\Gamma_2 \lesssim 0.5\,\Gamma_1$ considered here.

	Section~\ref{sec:multi-qubit} examines the dependence of the retained ergotropy on the number of qubits $N \geq 3$, using initial states drawn from the GHZ family that generalises $\ket{\Phi^+}$.

\section{Multi-qubit case: Dependence on battery size}
	\label{sec:multi-qubit}
	
	The two-qubit analysis of Sec.~\ref{sec:N2} showed that the protection
	mechanism remains effective across a range of initial preparations and in
	the presence of direct qubit--qubit interactions. A separate question
	concerns its dependence on the number of battery cells $N$. While
	collective advantages have been extensively studied during the charging
	stage of quantum batteries~\cite{Campaioli2017,Ferraro2018,Crescente2020,JadElAllati2026},
	their manifestation during the storage phase, where the dynamics is
	dominated by dissipation, remains less explored. In this section, we
	investigate how the cavity-mediated protection mechanism scales with
	system size.
	
	We consider batteries containing $N \in \{2,3,4,5\}$ qubits, initially
	prepared in the GHZ state
	\begin{equation}
		\ket{\Psi_{\mathrm{GHZ}}^{(N)}}
		= \frac{1}{\sqrt{2}}\bigl(\ket{e}^{\otimes N} + \ket{g}^{\otimes N}\bigr),
		\qquad
		\label{eq:GHZ_state}
	\end{equation}
	For $N=2$, Eq.~\eqref{eq:GHZ_state} reduces to the Bell state
	$\ket{\Phi^+}$ considered in Sec.~\ref{sec:N2}, providing a natural
	extension of the same benchmark to larger batteries. The GHZ state
	combines maximal coherence between the fully excited and fully unexcited
	sectors with a simple and well-defined scaling in system size, making it
	a convenient probe of collective effects in the storage dynamics.
	
	To connect with realistic implementations, we additionally include an
	all-to-all Heisenberg interaction between the qubits,
	\begin{equation}
		\hat H_{qq}^{(\mathrm{Heis})}
		= g\sum_{1 \le i < j \le N}
		\bigl(\hat\sigma_x^{(i)}\hat\sigma_x^{(j)}
		+ \hat\sigma_y^{(i)}\hat\sigma_y^{(j)}
		+ \hat\sigma_z^{(i)}\hat\sigma_z^{(j)}\bigr),
		\label{eq:Heis_all_to_all}
	\end{equation}
	Interactions of this form arise naturally in cavity-QED platforms through
	cavity-mediated exchange processes between qubits coupled to a common
	mode. Moreover, the Heisenberg Hamiltonian constitutes a special case of
	the broader family of collective XYZ spin models that have recently been
	engineered experimentally in optical-cavity settings~\cite{Luo2025XYZ}.
	The parameter regime considered here is therefore compatible with current
	cavity-QED architectures. Each component $\ket{e}^{\otimes N}$ and $\ket{g}^{\otimes N}$ is annihilated by $\hat\sigma_x^{(i)}\hat\sigma_x^{(j)} + \hat\sigma_y^{(i)}\hat\sigma_y^{(j)} = 2(\hat\sigma_+^{(i)}\hat\sigma_-^{(j)} + \hat\sigma_-^{(i)}\hat\sigma_+^{(j)})$ and is an eigenstate of $\hat\sigma_z^{(i)}\hat\sigma_z^{(j)}$ with eigenvalue $+1$, so that
	\begin{equation}
		\hat H_{qq}^{(\mathrm{Heis})}\,\ket{\Psi_{\rm GHZ}^{(N)}}
		= g\,\frac{N(N-1)}{2}\,\ket{\Psi_{\rm GHZ}^{(N)}}.
		\label{eq:GHZ_eigen}
	\end{equation}
	The Heisenberg interaction therefore contributes only an overall phase to the GHZ state and does not induce transitions within the qubit subsystem. As a result, the dependence on (N) discussed below primarily reflects the cavity-mediated protection mechanism rather than the direct qubit--qubit interaction.
	
	We work at the intermediate temperature $\beta\omega = 1$ ($\bar n \approx 0.58$): high enough that thermal noise injects a non-negligible \begin{figure*}[t]
		\centering
		\includegraphics[width=\textwidth]{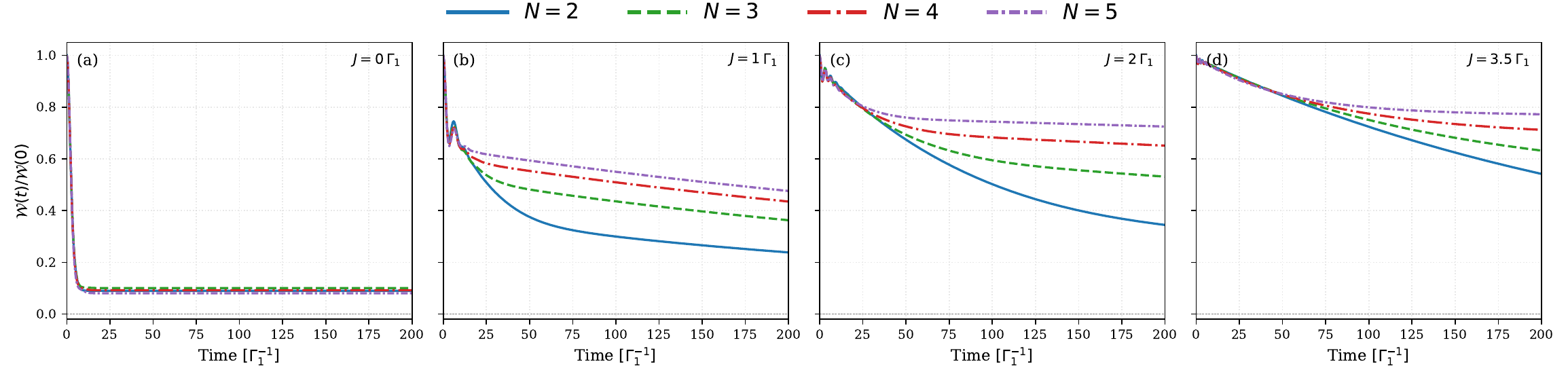}
		\caption{Normalised ergotropy $\mathcal{W}(t)/\mathcal{W}(0)$ of the multi-qubit battery prepared in the GHZ state of Eq.~\eqref{eq:GHZ_state}, in the presence of an all-to-all Heisenberg coupling (Eq.~\eqref{eq:Heis_all_to_all}, $g = 0.5\,\Gamma_1$), for $N=2$ (blue solid), $N=3$ (green dashed), $N=4$ (red dash-dotted), and $N=5$ (purple dash-double-dotted). Panels (a)--(d) correspond to $J/\Gamma_1 = 0,\,1,\,2,\,3.5$ respectively.  Parameters: $\kappa = 0.24\,\Gamma_1$, $\Gamma_2 = 0$, resonant configuration $\omega_0 = \omega_1 = \omega_2 \equiv \omega$, $\beta\omega = 1$ ($\bar n \approx 0.58$).}
		\label{fig:GHZ_Nscan}
	\end{figure*}
	 amount  of incoherent excitations through $\mathcal{D}[\hat a_n^\dagger]$, yet low enough that the cavity-mediated protection still operates on the same timescale as the dissipative drive.

	 All remaining parameters ($\omega$, $\kappa = 0.24\,\Gamma_1$, $\Gamma_2 = 0$) are identical to those of Secs.~\ref{sec:N1} and~\ref{sec:N2}. The time window is extended to $t\Gamma_1 \in [0, 200]$, four times longer than in the previous sections, to characterise the long-time retention of ergotropy beyond the initial transient regime. We consider four representative values of the inter-cavity coupling, $J/\Gamma_1 \in \{0,\,1,\,2,\,3.5\}$,

	  chosen to span the range of behaviours observed in the previous sections, from the uncoupled configuration to the regime of strong inter-cavity coupling.

	Figure~\ref{fig:GHZ_Nscan} displays the normalized ergotropy
	$\mathcal{W}(t)/\mathcal{W}(0)$ for the GHZ initial state of
	Eq.~\eqref{eq:GHZ_state} as a function of time for $N = 2,3,4,5$ and four
	values of the inter-cavity coupling $J$. In the absence of inter-cavity
	coupling [Fig.~\ref{fig:GHZ_Nscan}(a)], all system sizes exhibit nearly
	identical initial decay rates, with the ergotropy rapidly decreasing
	within $t\Gamma_1 \lesssim 5$. At longer times, however, a small size
	dependence becomes visible, where the case $N=2$ approaches a slightly
	lower residual ergotropy than the larger systems, whereas the $N=3,4,5$
	curves remain almost indistinguishable.
	
	For all finite inter-cavity couplings [Figs.~\ref{fig:GHZ_Nscan}(b)--(d)],
	the dynamics exhibit a clear size dependence that is absent at $J=0$.
	Following the initial transient oscillations, the ergotropy curves
	separate and remain ordered according to the number of qubits, with
	larger systems consistently retaining a greater fraction of their initial
	ergotropy. In particular, $N=5$ displays the highest long-time ergotropy,
	whereas $N=2$ exhibits the strongest degradation. This ordering persists
	throughout the simulated time window and becomes more pronounced as the
	system approaches its long-time regime.
	
	The figure also shows that increasing the inter-cavity coupling enhances
	ergotropy preservation for all system sizes. As $J$ is increased from
	$\Gamma_1$ to $3.5\,\Gamma_1$, the long-time ergotropy rises
	substantially and the decay becomes progressively suppressed. These
	results indicate that the auxiliary cavity not only mitigates the loss of
	ergotropy but does so more effectively for larger qubit ensembles.
	
	The enhanced ergotropy retention observed for larger $N$ can be
	understood from the collective nature of the qubit--cavity interaction.
	Within the symmetric Dicke manifold, the collective lowering operator
	$\hat S_- = \sum_{i=1}^{N}\hat\sigma_-^{(i)}$ acts on the GHZ state
	according to
	\begin{equation}
		\hat S_-\,\ket{\Psi_{\rm GHZ}^{(N)}}
		= \frac{\sqrt{N}}{\sqrt{2}}\,
		\ket{J = \tfrac{N}{2},\,M = \tfrac{N}{2} - 1},
		\label{eq:Sminus_GHZ}
	\end{equation}
	which reflects the well-known $\sqrt{N}$ enhancement of the collective
	Tavis--Cummings coupling. As a result, increasing the number of qubits
	strengthens the coherent exchange of excitations between the qubit
	ensemble and the cavity subsystem. In the presence of the auxiliary
	cavity, this enhanced collective coupling allows a larger fraction of the
	excitation to remain stored within the protected qubit--cavity sector
	rather than being irreversibly lost through the dissipative channel.
	Consequently, the long-time ergotropy increases with $N$, consistent with
	the ordering observed in Fig.~\ref{fig:GHZ_Nscan}.
	
	This collective enhancement is reminiscent of the many-body advantages
	reported during the charging stage of quantum
	batteries~\cite{Ferraro2018,Crescente2020,JadElAllati2026}. Here, however,
	its role is different: rather than accelerating energy storage, it
	improves the preservation of stored work against dissipation.
	
	The influence of the system size remains visible throughout the explored
	coupling range, although its manifestation depends on the decay
	timescale. Since all curves are normalized by their initial ergotropy,
	they begin from the same value and separate only as dissipation
	progressively degrades the stored work. For moderate couplings, where the
	residual decay is still appreciable within the simulated time window, the
	ordering with $N$ becomes clearly visible at relatively early times. As
	the inter-cavity coupling is increased, ergotropy preservation improves
	substantially and the decay is strongly suppressed. Consequently, the
	curves remain closer together over the same observation interval, not
	because the size dependence disappears, but because the slower dynamics
	provide less opportunity for differences between system sizes to
	accumulate. The ordering $\mathcal{W}_{N+1}(t) > \mathcal{W}_N(t)$
	nevertheless persists at long times for all $J > 0$ considered, indicating
	that the collective enhancement associated with larger $N$ survives even
	in the strongly protected regime.

	\section{Experimental feasibility}
	\label{sec:experimental}
	
	The cavity-based architecture analysed in this work has a natural implementation in circuit quantum electrodynamics (cQED), in which transmon qubits embedded in coplanar-waveguide resonators provide a realisation of the Tavis--Cummings coupling, and in which coherent photon exchange between two nearby resonators is engineered through capacitive coupling, a tunable coupler, or a shared bus mode~\cite{Blais2021cQED}. We briefly translate the dimensionless parameters used here into laboratory units.
	
	In modern cQED platforms the cavity frequency is typically $\omega_n/2\pi \sim 5$--$10$~GHz, and the qubit transition frequency $\omega_0/2\pi$ can be tuned in resonance with the cavity through a transmon SQUID loop~\cite{Blais2021cQED,Hu2022}. The decay rate of the lossy cavity $C_1$ is design-tunable in the range $\Gamma_1/2\pi \sim 1$--$100$~MHz, while a high-quality auxiliary cavity $C_2$ routinely achieves $\Gamma_2/2\pi \lesssim 10$~kHz, placing realistic devices in the range $\Gamma_2/\Gamma_1 \sim 10^{-5}$--$10^{-2}$. As shown by the robustness analysis of Sec.~\ref{sec:N2} (Figs.~\ref{fig:N2_Phi} and~\ref{fig:N2_ee}), the retained ergotropy is essentially independent of $\Gamma_2$ across this range and up to $\Gamma_2 \lesssim 0.5\,\Gamma_1$, so the protection does not require a lossless auxiliary cavity~\cite{Blais2021cQED}. The fixed value $\kappa = 0.24\,\Gamma_1$ used here places the bare $J = 0$ qubit dynamics in the weak-coupling Markovian regime: for $\Gamma_1/2\pi \sim 10$~MHz this corresponds to $\kappa/2\pi \sim 2.4$~MHz, set by the qubit--cavity coupling capacitance and the resonator geometry while keeping the qubit on resonance with the cavity ($\omega_0 = \omega_1$). The inter-cavity sweep $J/\Gamma_1 \in [0,5]$ employed in Secs.~\ref{sec:N1}--\ref{sec:multi-qubit} then translates to $J/2\pi \in [0,50]$~MHz, comparable to the photon-exchange couplings realised between adjacent cQED resonators~\cite{Blais2021cQED}. The local form of the dissipator in Eq.~\eqref{eq:master} is consistent with these numbers: the largest value $J = 5\,\Gamma_1$ corresponds to $J/\omega_n \sim 7 \times 10^{-3}$ for $\omega_n/2\pi = 7$~GHz and $\Gamma_1/2\pi = 10$~MHz, well within the validity regime of the local approach~\cite{Hofer2017,Cattaneo2019,DeChiara2018}.
	
	The dimensionless inverse temperature $\beta\omega$ sets the mean thermal occupation $\bar n = (e^{\beta\omega}-1)^{-1}$ in the cavities. At the base temperature $T \approx 15$--$20$~mK of a dilution refrigerator and $\omega/2\pi \approx 7$~GHz, the equilibrium $\bar n$ is below $10^{-7}$; the residual $\bar n \sim 10^{-3}$ typically reported in cQED is dominated by non-equilibrium stray photons rather than by the equilibrium Bose--Einstein occupation~\cite{Blais2021cQED}. Either way, this is effectively the zero-temperature limit considered in Ref.~\cite{Man2015}, and our predictions reduce to the long-lived plateaus of Fig.~\ref{fig:N1} (left column). The intermediate-temperature regime $\beta\omega \approx 1$ ($\bar n \approx 0.58$) at which most of our analysis was performed can be reached by intentionally injecting microwave noise into $C_1$ to set a controlled thermal photon population while keeping the qubits cold, or by lowering the cavity frequency into the sub-GHz range with a lumped-element resonator, which moves the same $\bar n$ window into the standard cryogenic-bath regime~\cite{Blais2021cQED}.
	
	The single-qubit preparations of Eqs.~\eqref{eq:N1_state1}--\eqref{eq:N1_state2} and the Bell state $\ket{\Phi^+}$ of Sec.~\ref{sec:N2} are standard targets in cQED~\cite{Hu2022}; GHZ states for $N \le 5$ have been prepared and characterised with full tomography on superconducting platforms~\cite{Barends2014}, and the readout of multi-qubit ergotropy has been demonstrated in the star-topology NMR architecture of Ref.~\cite{JoshiMahesh2022}. The organic-microcavity setup of Ref.~\cite{Quach2022} provides a complementary platform in which the Tavis--Cummings coupling between many two-level emitters and a single optical mode is realised. The parameter window used here, $\kappa/\Gamma_1 = 0.24$, $J/\Gamma_1 \in [0,5]$, $\Gamma_2/\Gamma_1 \lesssim 10^{-2}$, and $\beta\omega \in [0.3, 1.75]$, falls within the operational range of present-day cQED hardware~\cite{Man2015,Blais2021cQED,Hu2022}.

	\section{Conclusion and outlook}
	\label{sec:conclusion}
	
	We have investigated the cavity-mediated suppression of self-discharging
	in open quantum batteries operating at finite temperature, extending the
	analysis from single-qubit batteries to multi-qubit systems. The model
	consists of qubits coupled to a lossy cavity that is coherently connected
	to an auxiliary cavity acting as a structured environment, and the
	battery performance is quantified through the ergotropy of the reduced
	qubit state.
	
	A key outcome of our analysis is that coupling the battery cavity to an
	auxiliary resonator significantly extends the lifetime of stored work.
	The resulting protection survives at finite temperature and remains
	effective across a wide range of system parameters, underscoring the
	robustness of the proposed architecture.
	
	For a single-qubit battery, we find that coherent superposition states
	retain a larger fraction of their initial ergotropy than incoherent, yet
	fully excited preparations. Extending the analysis to two-qubit
	batteries, Bell-state preparations exhibit enhanced long-time ergotropy
	retention compared with the fully excited state, while the inclusion of
	Heisenberg-type qubit--qubit interactions leaves this protection
	mechanism largely unchanged within the parameter regime considered.
	
	The multi-qubit analysis with GHZ-charged batteries and all-to-all
	Heisenberg interactions revealed that the cavity-assisted protection
	mechanism becomes increasingly effective as the number of qubits grows.
	Larger batteries retain a greater fraction of their initial ergotropy
	throughout the evolution, pointing to a beneficial interplay between
	collective many-body effects and the structured cavity environment.
	
	Several natural directions for future work arise from the present study.
	First, while the auxiliary cavity already incorporates dissipation,
	exploring more general non-idealities such as asymmetric loss rates or
	additional noise sources could further clarify the robustness of the
	observed size-dependent protection.
	
	Second, the present multi-qubit analysis has been restricted to GHZ-type
	initial states. Extending the study to other relevant families of
	many-body states, including Dicke, $W$, and partially separable coherent
	product states, would help determine whether the observed monotonic
	enhancement with system size is a generic feature of collective coupling
	or specific to GHZ-type correlations.
	
	The parameter regimes considered here are accessible with current
	circuit-QED technology, where single- and few-qubit implementations are
	already well established, and multi-qubit platforms with collective
	readout are rapidly advancing toward larger-scale realizations.

	\begin{acknowledgments}
		The authors gratefully acknowledge that this research was supported through computational resources of HPC-MARWAN (\href{http://hpc.marwan.ma}{hpc.marwan.ma}) provided by the National Center for Scientific and Technical Research (CNRST), Rabat, Morocco.
	\end{acknowledgments}
	
	\appendix
	
	\section{The dark mode of Eq.~\eqref{eq:darkmode}}
	\label{app:darkmode}
	
	In this appendix we derive the single-excitation dark mode $|\psi_-\rangle$ of
	Eq.~(\ref{eq:darkmode}) and establish the two properties invoked in the main
	text: that it is annihilated by the lossy-cavity operator $\hat{a}_1$, and that
	the corresponding rank-one density operator is a fixed point of the master
	equation~(\ref{eq:master}) in the zero-temperature limit. The derivation
	proceeds by restricting the coherent Hamiltonian to the single-excitation
	sector of the qubit$+C_1+C_2$ system, diagonalising the resulting $3\times3$
	matrix, and identifying the zero-energy eigenvector.
	
	We work in the rotating frame at the common frequency
	$\omega = \omega_0 = \omega_1 = \omega_2$. The total Hamiltonian restricted to
	the single-excitation sector
	\begin{equation}
		\mathcal{H}_1 = \mathrm{span}\bigl\{\,
		|e,0,0\rangle,\; |g,1,0\rangle,\; |g,0,1\rangle \,\bigr\}
	\end{equation}
	reads, in the ordered basis above,
	\begin{equation}
		\hat{H}\big|_{\mathcal{H}_1} =
		\begin{pmatrix}
			0 & \kappa & 0 \\
			\kappa & 0 & J \\
			0 & J & 0
		\end{pmatrix}.
		\label{eq:H1matrix}
	\end{equation}
		Diagonalising $\hat{H}\big|_{\mathcal{H}_1}$, the characteristic equation
	$\det(\hat{H}\big|_{\mathcal{H}_1} - \lambda I) = 0$ gives the eigenvalues
	$\lambda \in \{-\Omega,\, 0,\, +\Omega\}$, where $\Omega = \sqrt{J^2+\kappa^2}$.
	To obtain the dark state, we consider the zero eigenvalue $\lambda = 0$.
	Writing the corresponding eigenvector as $v = (v_1, v_2, v_3)^{\mathsf{T}}$, the
	equation $\hat{H}\big|_{\mathcal{H}_1}\, v = 0$ implies $v_2 = 0$ and
	$\kappa v_1 + J v_3 = 0$, yielding $v \propto (J,\,0,\,-\kappa)$. After
	normalisation,
	\begin{equation}
		|\psi_-\rangle = \frac{1}{\sqrt{J^2+\kappa^2}}
		\bigl(\,J\,|e,0,0\rangle - \kappa\,|g,0,1\rangle\,\bigr),
	\end{equation}
	which is Eq.~(\ref{eq:darkmode}). Both surviving components have $\ket{0}_{C_1}$, so $\hat a_1\ket{\psi_-} = 0$ and $\hat a_1^\dagger\hat a_1\ket{\psi_-} = 0$.
	
	The density operator
	\[
	\rho_- = |\psi_-\rangle\langle\psi_-|
	\]
	is therefore a stationary state of the master equation (3) in the zero-temperature limit ($\bar n=0$). Indeed, the coherent contribution vanishes since $|\psi_-\rangle$ is an eigenstate of $\hat H$ with eigenvalue zero,
	\[
	\hat H|\psi_-\rangle = 0
	\quad \Rightarrow \quad
	[\hat H,\rho_-]=0,
	\]
	while the dissipative term vanishes because
	\[
	\hat a_1|\psi_-\rangle = 0,
	\]
	which implies
	\[
	D[\hat a_1]\rho_- = 0.
	\]
	Hence,
	\[
	\dot{\rho}_- = 0,
	\]
	and $\rho_-$ is a stationary state of the dynamics.
	
\begin{figure}[h]
		\centering
		\includegraphics[width=\columnwidth]{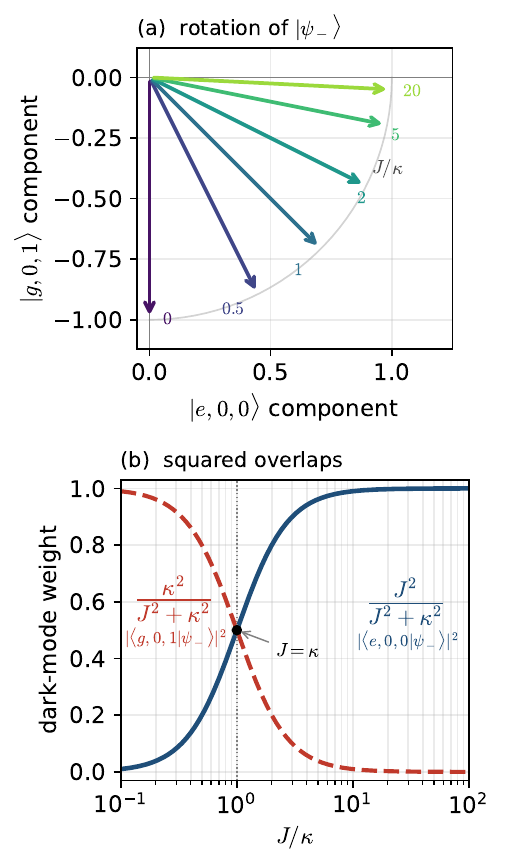}
		\caption{Composition of the single-excitation dark mode $|\psi_-\rangle$ of
			Eq.~(\ref{eq:darkmode}) as a function of $J/\kappa$. (a) $|\psi_-\rangle$ shown as
			a unit vector in the loss-free plane $\{|e,0,0\rangle,\,|g,0,1\rangle\}$ for
			representative values of $J/\kappa$ (colour-coded); the eigenvector rotates from
			$-|g,0,1\rangle$ at $J/\kappa=0$ toward $|e,0,0\rangle$ as $J/\kappa\to\infty$,
			with no overshoot. (b) Squared overlaps of $|\psi_-\rangle$ with the two basis
			states, $J^2/(J^2+\kappa^2)$ (blue solid) and $\kappa^2/(J^2+\kappa^2)$ (red
			dashed); the weights cross at $J=\kappa$. The blue curve equals the asymptotic
			coherence-retention fraction quoted in the main text.}
		\label{fig:saturation}
	\end{figure}

	The dependence of $\ket{\psi_-}$ on $J/\kappa$ obtained above is visualised in Fig.~\ref{fig:saturation}.
	
	The composition of the dark mode $|\psi_-\rangle$ as a function of the ratio
	$J/\kappa$ is shown in Fig.~\ref{fig:saturation}. Panel~(a) plots $|\psi_-\rangle$
	as a unit vector in the loss-free plane $\{|e,0,0\rangle,\,|g,0,1\rangle\}$; the
	third component, on the lossy state $|g,1,0\rangle$, is identically zero, which is
	what makes $|\psi_-\rangle$ annihilated by $\hat{a}_1$. At $J/\kappa = 0$ the dark
	mode coincides with $-|g,0,1\rangle$ (excitation entirely in the auxiliary cavity,
	no qubit weight), and as $J/\kappa$ grows it rotates monotonically toward
	$|e,0,0\rangle$ without overshooting. Panel~(b) makes this quantitative by plotting
	the squared overlaps $|\langle e,0,0|\psi_-\rangle|^2 = J^2/(J^2+\kappa^2)$ (blue)
	and $|\langle g,0,1|\psi_-\rangle|^2 = \kappa^2/(J^2+\kappa^2)$ (red), which sum to
	unity and cross at $J=\kappa$. The blue curve is the weight of the dark mode on the
	energy-carrying qubit state, and coincides with the asymptotic coherence-retention
	fraction $J^2/(J^2+\kappa^2)$ quoted in Sec.~\ref{sec:N1}: as $J/\kappa$
	increases, a growing fraction of the qubit excitation is stored in the loss-free
	protected mode rather than in the channel exposed to cavity loss, which is the
	origin of the cavity-mediated protection.
	
	\section{Dependence on $\Gamma_2$ for fully excited preparation \texorpdfstring{$\ket{ee}$}{|ee>}}
	\label{app:ee}
	
	We complement Sec.~\ref{sec:N2} by presenting the corresponding
	results for the diagonal initial preparation $\rho_B(0) = \ketbra{ee}{ee}$. The
	reference time is chosen as $t_f\Gamma_1 = 7.53$, corresponding to the time at
	which the uncoupled case ($J = 0$) approaches within $1\%$ of its asymptotic
	ergotropy value. This reference time is shorter than for the Bell state
	$\ket{\Phi^+}$, reflecting the faster self-discharge of the fully excited
	preparation in the absence of cavity-mediated protection.
	
	Compared with the Bell-state results, two qualitative differences can be
	identified. First, at the longer evaluation times $5t_f$ and $10t_f$, the
	contours of constant $\mathcal{W}(t_f)/\mathcal{W}(0)$ exhibit a noticeable
	upward curvature, indicating an increased sensitivity of the retained ergotropy
	to the auxiliary-cavity dissipation rate $\Gamma_2$. By contrast, the
	corresponding contours for $\ket{\Phi^+}$ remain nearly vertical over the same
	parameter range, demonstrating a weaker dependence on $\Gamma_2$.
	
	Second, for small values of the inter-cavity coupling $J$, the fully excited
	state retains less ergotropy than the Bell state at the same values of
	$(J,\Gamma_2)$, particularly at the longest evaluation time $10t_f$.
	
	\begin{figure}[h]
		\centering
		\includegraphics[width=\columnwidth]{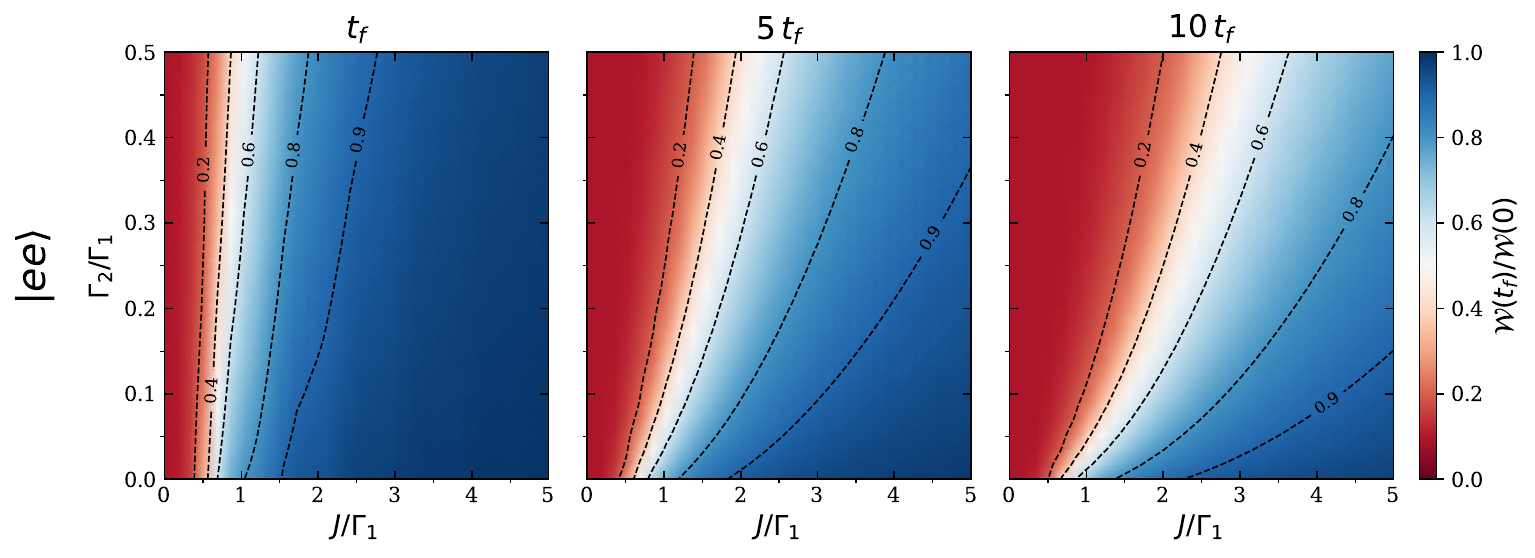}
		\caption{Retained ergotropy $\mathcal{W}(t_f)/\mathcal{W}(0)$ of the diagonal preparation $\ket{ee}$ in the plane of $J/\Gamma_1$ and $\Gamma_2/\Gamma_1$, at $\beta\omega = 1$. Panels show $t_f$, $5\,t_f$, and $10\,t_f$, with $t_f\Gamma_1 = 7.53$. Dashed contours as in Fig.~\ref{fig:N2_Phi}. Parameters: $\omega = 1$, $\kappa = 0.24\,\Gamma_1$, $g = 0.1\,\Gamma_1$.}
		\label{fig:N2_ee}
	\end{figure}

\end{document}